\documentclass[sigconf,nonacm,screen,10pt]{acmart}
\AtBeginDocument{%
  }

\usepackage{appendix}

\usepackage{graphicx}
\usepackage{microtype}
\usepackage{xspace}
\usepackage{subcaption}
\usepackage{wrapfig}
\usepackage{algorithm}
\usepackage{amsthm}
\usepackage{verbatim}
\usepackage{mathtools}
\usepackage{soul}
\usepackage{enumitem}
\usepackage{bbm}
\usepackage{multirow}
\usepackage{algpseudocode}
\usepackage{amsmath}

\usepackage[font=small]{caption}

\begin{document}

\title{Learning-Based Reconstruction Attacks on Coordinate-Obfuscated Point Clouds}


\author{Mohammad Waquas Usmani}
\email{mohammadwaqu@umass.edu}
\affiliation{%
  \institution{University of Massachusetts Amherst}
  \state{Massachusetts}
  \country{USA}
  \thanks{© 2026 The Authors. This is the author's version of the work. The definitive Version of Record will appear in \textit{Second Workshop on Enhancing Security, Privacy, and Trust in Extended Reality (XR Security '26)}, October 26--30, 2026, Austin, TX, USA. DOI: \url{https://doi.org/10.1145/3842203.3844590}.}
}
\author{Susmit Shannigrahi}
\email{sshannigrahi@tntech.edu}
\affiliation{%
  \institution{Tennessee Technological University}
  \state{Tennessee}
  \country{USA}
}
\author{Michael Zink}
\email{zink@ecs.umass.edu}
\affiliation{%
  \institution{University of Massachusetts Amherst}
  \state{Massachusetts}
  \country{USA}
}

\renewcommand{\shortauthors}{Usmani et al.}

\begin{abstract}

Volumetric video based on point cloud representations enables immersive virtual and augmented reality applications but introduces significant challenges for efficient and secure content delivery. Prior work proposed a selective coordinate encryption framework for point clouds that encrypts only a subset of coordinates, reducing computational costs while visually degrading unauthorized content. However, it remains unclear whether the remaining unencrypted information is sufficient to enable content reconstruction.

In this paper, we evaluate the robustness of selective coordinate encryption against machine learning-based reconstruction attacks. We consider an attacker with access to selectively encrypted point clouds attempting to recover encrypted coordinates without decryption by exploiting spatial and geometric correlations in the unencrypted data. We evaluate PointNet and Random Forest models under two encryption granularities: \texttt{X}, where all $X$ coordinates are encrypted, and \texttt{2X}, where every second $X$ coordinate is encrypted


Our results show that reconstructing fully encrypted $X$ coordinates remains challenging, whereas the \texttt{2X} scheme leaks sufficient information through neighboring coordinates to enable accurate reconstruction. These findings demonstrate that the security of selective coordinate encryption depends strongly on encryption granularity.



\end{abstract}

\keywords{Machine Learning, Point Clouds, Volumetric Video, Selective Encryption, Coordinate Encryption, Reconstruction Attacks, Deep Learning, Digital Rights Management}



\maketitle

\section{Introduction} \label{sec:introduction}

Volumetric video and point cloud content are increasingly important for immersive virtual reality, augmented reality, and collaborative 3D applications. Unlike traditional 2D video, point clouds consist of large collections of spatial coordinates and associated attributes describing a 3D scene. Their large data volume introduces substantial computational and bandwidth overheads, increasing latency and motion-to-photon delay, contributing to motion sickness~\cite{10.1145/3386290.3396933}.


Protecting volumetric content from unauthorized access and redistribution is another challenge. While transport-layer encryption secures communication channels, it does not provide efficient protection tailored for point cloud streaming. Although several studies have investigated encryption of individual point clouds~\cite{10226134}, comparatively little attention has been given to securing volumetric video streams.


To address this gap, our prior work proposed a selective coordinate encryption framework for point cloud Digital Rights Management (DRM)~\cite{10.1145/3704413.3765298}. Rather than encrypting entire point clouds, the framework encrypts only selected coordinates, reducing computational overhead while degrading the visual quality of unauthorized content. However, because only a subset of the point cloud information is encrypted, an important security question emerges: \textit{can an attacker reconstruct the missing encrypted coordinates using the remaining unencrypted, visible information?}


Modern machine learning (ML) models can effectively learn spatial and geometric relationships in point clouds for tasks such as classification, segmentation, reconstruction, and completion~\cite{7353481,Qi_2017_CVPR,3295222.3295263,9173706,Yang_2018_CVPR,8491026,Xie_2021_CVPR,Yu_2018_CVPR,11366821}. Consequently, selectively encrypted point clouds may leak sufficient information through visible coordinates and attributes to enable recovery of protected coordinates without decryption.


In this work, we evaluate the robustness of selective coordinate encryption against ML-based reconstruction attacks. We consider an attacker with access to selectively encrypted point clouds who attempts to reconstruct encrypted coordinates by exploiting spatial and geometric relationships in the visible data. To this end, we evaluate PointNet~\cite{Qi_2017_CVPR} and Random Forest models using different combinations of spatial, geometric, and color features under two encryption granularities: \texttt{X}, where all $X$ coordinates are encrypted, and \texttt{2X}, where every second $X$ coordinate is encrypted.



For the \texttt{X} scheme, reconstructing all encrypted $X$ coordinates remains challenging, resulting in relatively high prediction errors and noticeable geometric deviations. Feature importance analysis further indicates that the surface normal component $n_x$, followed by the visible $Y$ and $Z$ coordinates, provides the most useful information for reconstruction.


In contrast, the \texttt{2X} scheme leaks substantial information through neighboring visible $X$ coordinates, enabling significantly accurate reconstruction of point cloud geometry. These findings show that the security of selective coordinate encryption depends critically on encryption granularity and highlight the importance of carefully balancing computational efficiency with resistance to reconstruction attacks.



\section{Background and Related Work} \label{sec:background}



\textbf{3D Point Clouds:}  Point clouds represent 3D objects or scenes as collections of points with spatial coordinates \((x, y, z)\) and optional attributes  such as color or surface normals. They are commonly stored in the Polygon File Format (\texttt{.PLY}). A sequence of point cloud frames forms a volumetric video, enabling 6DoF immersive experiences.
\\
\textbf{Securing 3D Point Clouds:} Numerous studies have investigated protecting point clouds using chaotic mapping, holographic methods, permutation and rotation-based encryption~\cite{10226134}. While these approaches provide strong protection for individual point clouds, they often target standalone content rather than real-time volumetric video streaming.

Our prior work~\cite{10.1145/3704413.3765298} proposed a selective coordinate encryption framework for efficient point cloud video streaming. Rather than encrypting an entire point cloud, the framework selectively encrypts subsets of spatial coordinates while leaving the remaining coordinates, attributes, and metadata unencrypted, balancing visual distortion with computational efficiency. Encrypting only the $X$ coordinate reduced encryption and decryption times by 37\% and 46\%, respectively, compared to full point cloud encryption.


This framework utilizes Attribute-Based Encryption~\cite{4223236}, motivated by our prior streaming studies~\cite{10.1145/3712676.3714450,usmani2025securingimmersive360video}. Building upon this, ABE-VVS~\cite{usmani2026abevvsattributebasedencryptedvolumetric} integrated attribute-based selective coordinate encryption for volumetric video delivery, reducing server and cache CPU usage while improving client QoE.


However, the resilience of selective coordinate encryption against machine learning-based reconstruction attacks remains unexplored. In particular, it is unknown whether the remaining visible coordinates and attributes leak sufficient information to recover encrypted geometry. This work addresses that gap by evaluating reconstruction attacks using modern machine learning models.
\\
\textbf{Machine Learning-based Reconstruction:} Machine learning has been widely applied to point cloud understanding and geometric reconstruction. Comprehensive surveys of point cloud completion categorize existing approaches into traditional geometric methods and deep learning-based methods, including PointNet-based, GAN-based, Transformer-based, and point upsampling techniques~\cite{10433645}.

PointNet and PointNet++~\cite{Qi_2017_CVPR,3295222.3295263} demonstrated the ability to directly learn spatial and geometric features from unordered point clouds, enabling accurate classification, segmentation, and object recognition. Subsequently, learning-based methods such as FoldingNet~\cite{Yang_2018_CVPR}, PCN~\cite{8491026}, Xie et al.~\cite{Xie_2021_CVPR}, and PU-Net~\cite{Yu_2018_CVPR} extended these ideas to point cloud completion, reconstruction, and upsampling. More recently~\cite{11366821,doi:10.1142/S1793351X2644006X} proposed a Random Forest-based point cloud upsampling framework integrated with selective coordinate encryption.

Unlike conventional point cloud completion, which reconstructs missing geometry from incomplete observations, our work investigates a reconstruction attack in which selectively encrypted coordinates are inferred from the remaining unencrypted information within a protected point cloud. To quantify the information leakage, we evaluate both PointNet and Random Forest models for reconstructing encrypted coordinates from visible coordinates, neighborhood information, and other point attributes.

\section{Methodology}
\label{sec:methodology}
\subsection{Threat Model}

\begin{figure}[!ht]
\centering
  \includegraphics[width=1\columnwidth]{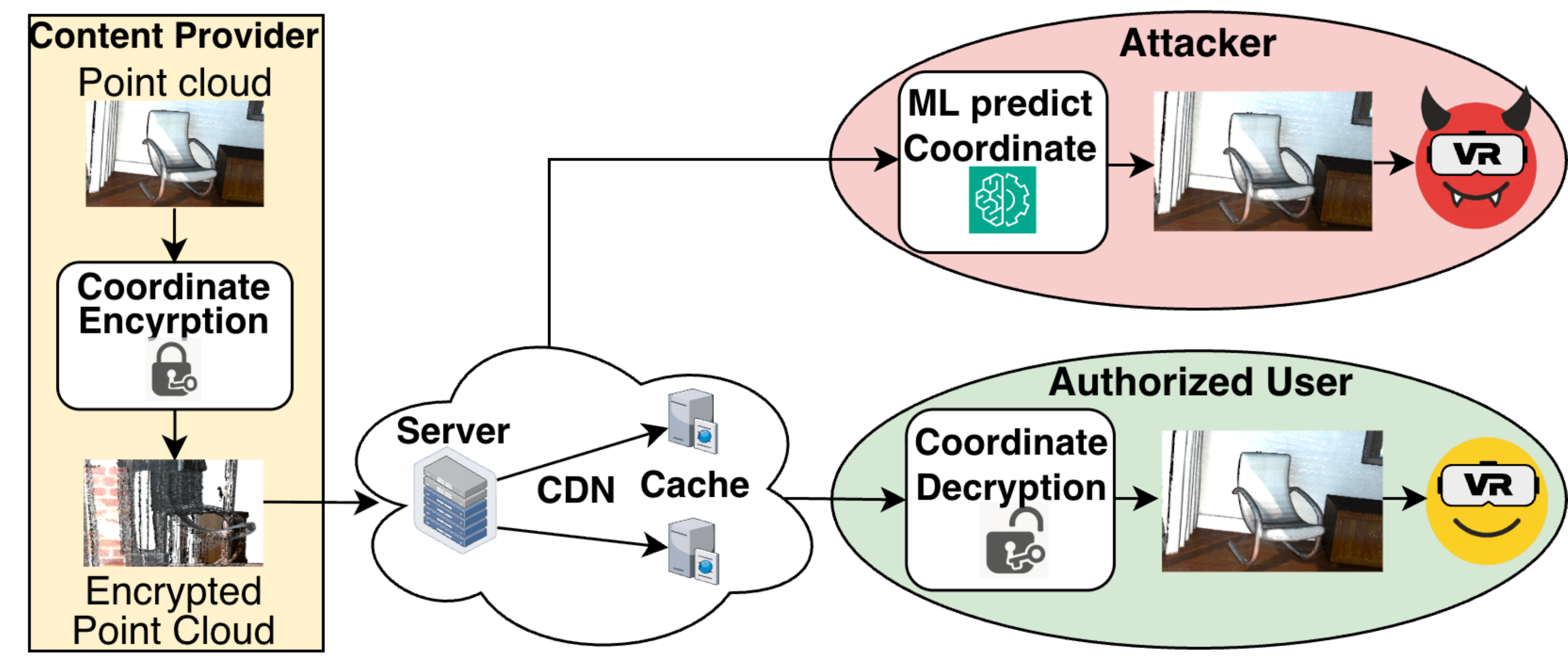} 
  \captionsetup{justification=centering}
  \caption{Illustration of the threat model.} 
  \label{fig:threat} 
\end{figure}

Figure~\ref{fig:threat} illustrates the threat model. A content provider selectively encrypts point cloud coordinates before distributing the content through a CDN-based streaming infrastructure. Authorized users possessing valid decryption keys can decrypt the protected coordinates and reconstruct the original point cloud for volumetric video playback.


We consider an unauthorized attacker with access to selectively encrypted point cloud frames but without the required decryption keys. Because only a subset of spatial coordinates is encrypted, while the remaining coordinates, surface normals, color attributes, and metadata remain visible, the attacker exploits this information leakage to reconstruct the protected geometry using ML techniques. We assume the attacker knows the encryption granularity and methodology and has access to publicly available point cloud datasets for training ML model capable of learning spatial and geometric relationships between visible and encrypted coordinates.




\emph{Unlike traditional cryptographic attacks aimed at breaking the encryption scheme itself, the objective of the attacker in our model is to infer or predict the encrypted coordinate values using machine learning-based reconstruction techniques.} By leveraging correlations among neighboring points, unencrypted coordinates, surface normals, and color attributes, the attacker attempts to reconstruct the original point cloud geometry and recover meaningful visual information without performing decryption. Our work evaluates whether selectively encrypted point clouds leak sufficient information to enable such reconstruction attacks.




\subsection{Data Preparation}
\label{subsec:data-prep}

Before training, each point cloud frame is independently normalized to a unit sphere by centering it at its centroid and scaling by its maximum Euclidean distance from the centroid. The point cloud frames used in our dataset contain varying numbers of points. To provide a consistent input size for PointNet (Sect.~\ref{subsec:PointNet_Model}), all frames are truncated to 100,000 points, corresponding to the smallest point count in the dataset. In contrast, the Random Forest model (Sect.~\ref{subsec:RandomForest_Model}) operates on individual point-level feature vectors and therefore utilizes all available points from each frame.


We evaluate two granularities of coordinate encryption: \textbf{\texttt{X}}, where the $X$ coordinates of each point are encrypted, and \textbf{\texttt{2X}}, where the $X$ coordinate of every second point is encrypted, leaving the alternating $X$ coordinates visible.

For the \texttt{X} scenario, reconstruction models use the visible coordinates $(Y,Z)$ together with surface normals $N=(n_x,n_y,n_z)$, color attributes $Col=(r,g,b)$, and additional geometric descriptors including local surface curvature ($Curv$) and average neighborhood distance ($ND$), computed using the $k=16$ nearest neighbors.

For the \texttt{2X} scenario, the partially observed coordinate sequence takes the form $X_2=(x_1,0,x_3,0,\ldots)$. Models use the same features as the \texttt{X} scenario, with additional configurations incorporating neighboring coordinate features $PX$ and $NX$, representing the previous and next unencrypted $X$ coordinates in the point cloud ordering.

\subsection{PointNet Reconstruction Model}
\label{subsec:PointNet_Model}

To evaluate the feasibility of reconstructing selectively encrypted coordinates using deep learning, we employ a PointNet-based regression architecture~\cite{Qi_2017_CVPR} that learns local and global geometric features directly from point cloud data.


Shared MLP layers with dimensions $d_{in}\!\rightarrow\!64\!\rightarrow\!64\!\rightarrow\!64\!\rightarrow\!128\!\rightarrow\!2048$, where $d_{in}$ denotes the input feature dimensionality, together with batch normalization and ReLU activations, extract point-level feature representations. To capture global geometric context, Symmetric max- and mean-pooling aggregate the 2048-dimensional point features into a 4096-dimensional global representation, which is combined with 64-dimensional local features. The combined features are passed through regression layers of size $4160\!\rightarrow\!512\!\rightarrow\!256\!\rightarrow\!1$ to predict the encrypted $X$ coordinates. The model contains about 2.55 million trainable parameters.


The model is trained using Mean Absolute Error (MAE) loss and the Adam optimizer (learning rate = 0.001, batch size = 8) for up to 100 epochs with early stopping (patience = 15). The model with the lowest training loss is used for evaluation on the held-out test set.


\subsection{Random Forest Reconstruction Model} \label{subsec:RandomForest_Model}

Furthermore, we evaluate Random Forest regression~\cite{10.1023/A:1010933404324} as a traditional machine learning baseline. Random Forests construct an ensemble of decision trees and aggregate their predictions to model complex non-linear relationships between input features and encrypted coordinate values.


Unlike PointNet, which processes entire point cloud frames, Random Forest operates on individual point-level feature vectors, each consisting of an encrypted coordinate and its associated visible features. The model uses 100 decision trees with a maximum depth of 20 and bootstrap aggregation. Feature configurations vary with encryption granularity, and model is evaluated on same held-out test set as PointNet.



\section{Evaluation} \label{sec:evaluation-1}

\subsection{Point Cloud Dataset}
\label{subsec:pointcloud-dataset}

The evaluation was conducted using the Open3D~\cite{Open3D} Office and Living Room point cloud datasets. Each point cloud contains coordinates (\(X,Y,Z\)), surface normals (\(N\)), and color attributes (\(Col\)). These datasets were selected as they represent realistic indoor scenes with complex geometric structures.


For all experiments, a subset of frames was reserved as a held-out test set and excluded from training. The Office dataset was split into 49 training and 4 test frames, while Living Room dataset was split into 53 training and 4 test frames. To simulate the \texttt{X} and \texttt{2X} encryption schemes, and assuming the attacker knows the encryption granularity, encrypted $X$ coordinates were replaced with zero as placeholders to preserve the point cloud format. Reconstruction performance was evaluated exclusively on the held-out test frames.


\subsection{Evaluation Metrics}
\label{subsec:eval-metrics}


We evaluate reconstruction quality using following metrics:

\textbf{Point-to-Point Mean Absolute Error (MAE):} directly measures the absolute difference between reconstructed and original $X$ coordinates. Since point correspondence is preserved, MAE is computed using the original point ordering.

\textbf{Chamfer Distance (CD):} Measures geometric similarity between reconstructed and original point clouds by computing the average nearest-neighbor distance in both directions. Lower CD values indicate higher geometric similarity.


\textbf{Hausdorff Distance (HD):} Measures the maximum nearest-neighbor distance between two point clouds, capturing the largest geometric deviation. HD is therefore particularly sensitive to outliers and localized reconstruction errors. Lower HD values indicate smaller worst-case geometric deviations.


Since all point clouds are normalized, MAE, CD, and HD are reported in normalized coordinate units.

\begin{table}[H]
\centering
\scriptsize
\caption{Different feature configurations evaluated}
\label{tab:feature_models}
\begin{tabular}{|c|l||c|l|}
\hline
\multicolumn{2}{|c||}{\textbf{\texttt{X} Encryption}} &
\multicolumn{2}{c|}{\textbf{\texttt{2X} Encryption}} \\
\hline
Model & Features & Model & Features \\
\hline
M1 & $Y,Z$ + $N$ + $Col$ + $ND$ + $Curv$
   & M1 & $X_2,Y,Z$ + $N$ + $Col$ \\

M2 & $Y,Z$ + $N$ + $Col$ + $ND$
   & M2 & $X_2,Y,Z$ + $N$ \\

M3 & $Y,Z$ + $N$ + $Col$
   & M3 & $X_2,Y,Z$ \\

M4 & $Y,Z$ + $N$
   & M4 & $PX$ + $NX$ + $Y,Z$ + $N$ + $Col$ \\

M5 & $Y,Z$
   & M5 & $PX$ + $NX$ + $Y,Z$ + $N$ \\

   &
   & M6 & $PX$ + $NX$ + $Y,Z$ \\

   &
   & M7 & $PX$ + $NX$ \\
\hline
\end{tabular}
\end{table}

\subsection{Attack on \texttt{X} Encryption Granularity}
\label{subsec:X-results}

We first evaluate the \texttt{X} encryption scenario, where the attacker attempts to reconstruct the $X$ coordinate of every point using the remaining visible information. Table~\ref{tab:feature_models} summarizes the feature configurations used for the PointNet and Random Forest models. For comparison, we also include a Zeroed-\texttt{X} baseline, in which encrypted $X$ coordinates are set to zero and no reconstruction is performed, providing a lower bound for evaluating reconstruction performance.


\begin{figure*}[t]
\centering
\includegraphics[width=0.86\textwidth]{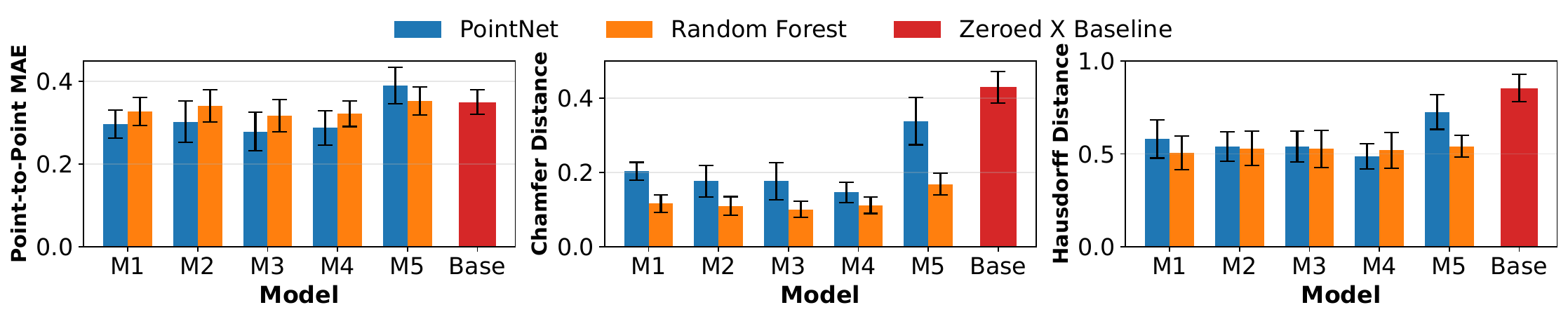}
\caption{Reconstruction performance for the \texttt{X} encryption attack across different models with 95\% confidence interval.}
\label{fig:x_attack}
\end{figure*}

\begin{figure*}[t]
\centering
\includegraphics[width=0.86\textwidth]{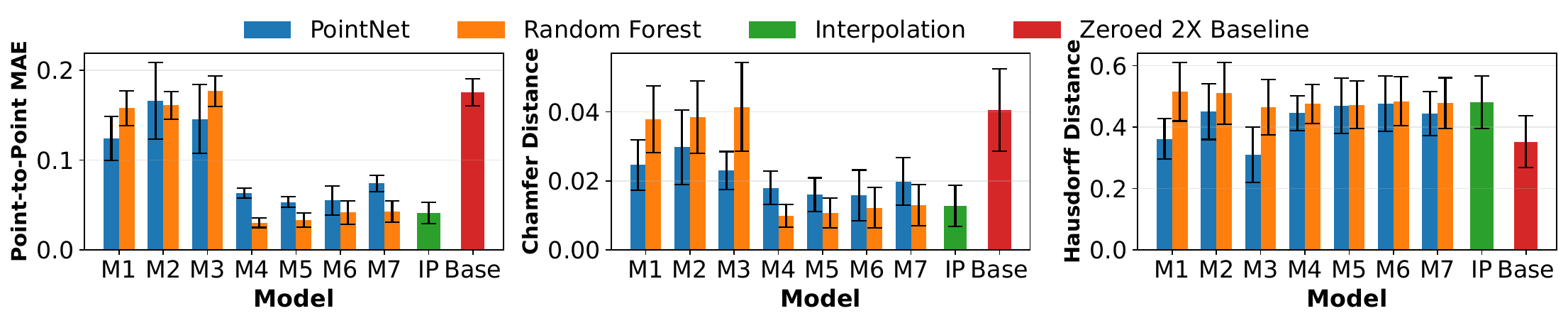}
\caption{Reconstruction performance for the \texttt{2X} encryption attack across different models with 95\% confidence interval.}
\label{fig:2x_attack}
\end{figure*}

\begin{figure*}[t]
\centering
\includegraphics[width=0.86\textwidth]{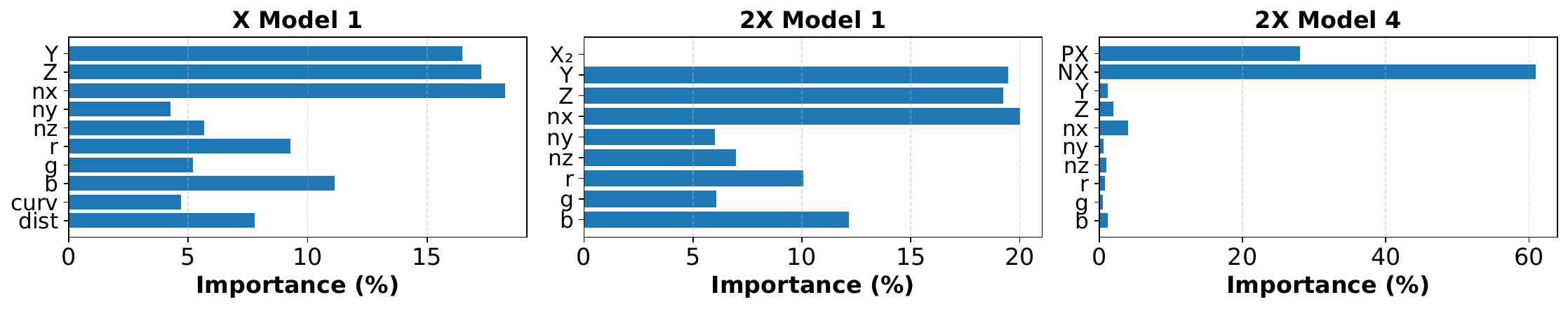}
\caption{Feature importance (\%) reported by Random Forest models for \texttt{X} and \texttt{2X} encryption granularities.}
\label{fig:feat_imp}
\end{figure*}


\textbf{Point-to-Point Reconstruction Accuracy:} Figure~\ref{fig:x_attack} (left) shows the average MAE of the reconstructed $X$ coordinates across the test set with 95\% confidence interval. PointNet and Random Forest achieve similar performance, with PointNet providing up to 4\% lower MAE across most configurations. Models M3 and M4 achieve the lowest error, while M1 and M2 perform slightly worse, indicating that local curvature and neighborhood distance provide little benefit. M5 exhibits the highest error, suggesting that surface normals contribute useful information beyond the visible $Y$ and $Z$ coordinates.

Compared to the Zeroed-\texttt{X} baseline (MAE = 35.0\%), both models reduce reconstruction error, with the best-performing PointNet M3, achieving an MAE of 27.9\%. However, the error remains substantial, indicating that accurate reconstruction of fully encrypted $X$ coordinates remains challenging.





\textbf{Geometric Reconstruction Accuracy:} Figure~\ref{fig:x_attack} (middle and right) presents the average CD and HD with 95\% confidence intervals. Random Forest consistently achieves lower CD and HD than PointNet, indicating better geometric reconstruction. Consistent with the MAE results, M1 and M2 show no benefit from local curvature and neighborhood distance. Instead, M4, which uses only visible coordinates and surface normals, achieves the lowest CD and HD, suggesting that surface normals provide useful geometric information, while their removal in M5 produces the largest degradation.


Compared to the Zeroed-\texttt{X} baseline (CD = 0.4295, HD = 0.8540), all models substantially improve geometric similarity, with Random Forest M4 achieving the lowest CD (0.11) and HD (0.52). Nevertheless, HD remains consistently higher than CD, reflecting localized reconstruction outliers due to its sensitivity to worst-case errors. Overall, although reconstruction improves geometric fidelity relative to the baseline, accurately recovering the original geometry remains challenging when all $X$ coordinates are encrypted.




\textbf{Impact of Feature Selection:} Figure~\ref{fig:feat_imp} (left) shows the Random Forest feature importance scores. The $Y$, $Z$, and $n_x$ features contribute most to inferring hidden $X$ coordinates, while the remaining normal components ($n_y$ and $n_z$), color attributes ($r,g,b$), local curvature, and neighborhood distance contribute considerably less.




\subsection{Attack on \texttt{2X} Encryption Granularity}
\label{subsec:2X-results}

In the \texttt{2X} encryption scenario, every second $X$ coordinate is hidden while neighboring $X$ coordinates remain visible. This direct coordinate leakage through adjacent points may enable interpolation-based reconstruction attacks. Table~\ref{tab:feature_models} summarizes the evaluated feature configurations.


Models M1–M3 evaluate whether half of the unencrypted $X$ coordinates ($X_2$), with combinations of other coordinates, surface normals, and color attributes, provide sufficient information to reconstruct encrypted $X$ values. Models M4–M7 additionally incorporate neighboring visible coordinates, $PX$ and $NX$, to evaluate impact of local coordinate leakage. The $Curv$ and $ND$ features are omitted because they provided little benefit in the \texttt{X} scenario. For comparison, we also include a Zeroed-\texttt{2X} baseline and a simple interpolation baseline.




{\textbf{Point-to-Point Reconstruction Accuracy:} Figure~\ref{fig:2x_attack} (left) presents the MAE results for the \texttt{2X} encryption scenario. The Zeroed-2X baseline already exhibits substantially lower error than the \texttt{X} baseline (17.5\% vs. 35.0\%), since only half of the $X$ coordinates are encrypted. Models M1–M3, which rely solely on the target point’s visible attributes, provide only modest improvements, with MAE values ranging from 12–18\%. In contrast, Models M4–M7, which incorporate the neighboring visible coordinates $PX$ and $NX$, achieve substantially lower errors. Simple interpolation using only $PX$ and $NX$ reduces the MAE to 4.1\%, outperforming M1–M3 and achieving performance comparable to Random Forest M7 (4.2\%), which also utilizes only $PX$ and $NX$. However, the best-performing configuration, Random Forest M4, further reduces the MAE to approximately 3\%, demonstrating that machine learning can exploit additional point attributes beyond simple interpolation to achieve more accurate reconstruction.



\textbf{Geometric Reconstruction Accuracy:} Figure~\ref{fig:2x_attack} (middle and right) presents the CD and HD results. Consistent with the MAE results, Models M1–M3 provide only modest improvements over the baseline (CD = 0.04). In contrast, Models M4–M7 substantially reduce the CD by incorporating $PX$ and $NX$. Simple interpolation achieves a CD of 0.013, whereas Random Forest M4 further reduces it to 0.009, showing that machine learning leverages additional point attributes beyond simple interpolation.


Interestingly, HD is higher than the baseline across most ML configurations and simple interpolation. Unlike CD, HD measures the maximum geometric deviation and is therefore highly sensitive to outliers. In the Zeroed-2X baseline, assigning all encrypted coordinates to $X=0$ produces a relatively uniform distortion. In contrast, reconstruction methods occasionally generate large errors for a small number of points, creating localized outliers that disproportionately increase the HD despite substantial improvements in MAE and CD.

\textbf{Impact of Feature Selection:} Figure~\ref{fig:feat_imp} (middle) shows the feature importance for M1. Similar to the \texttt{X} reconstruction scenario, the $Y$ and $Z$ coordinates together with the normal $n_x$ contribute most to reconstruction. The $X_2$ feature exhibits negligible importance because reconstruction targets correspond to points whose $X$ coordinates are encrypted and replaced with zeros, leaving no useful $X_2$ information available at the target location.


Figure~\ref{fig:feat_imp} (right) presents the feature importance for M4. Here, the neighboring visible coordinates $PX$ and $NX$ dominate all other features, while the contribution of remaining attributes becomes negligible. These results confirm that neighboring visible coordinates are the primary source of information leakage in the \texttt{2X} encryption scenario.


\section{Discussion and Conclusions} \label{sec:conclusion}

Our results show that reconstruction attacks depend strongly on encryption granularity. While reconstructing fully encrypted $X$ coordinates remains difficult, the \texttt{2X} scheme leaks sufficient information through neighboring visible coordinates to enable accurate reconstruction. Although simple interpolation provides a strong baseline, machine learning further exploits additional point attributes to improve reconstruction accuracy. Feature importance analysis further identifies neighboring coordinates and the $n_x$ surface normal component as the primary sources of information leakage. These findings suggest that protecting additional coordinate dimensions or surface normals may further improve resistance against reconstruction attacks.


Several directions remain for future work. More advanced reconstruction models, such as PointNet++~\cite{3295222.3295263}, larger datasets, and object-specific datasets may further improve attack performance. Alternative attacks based on point cloud upsampling~\cite{11366821} may also be effective against the \texttt{2X} scheme.
Evaluating additional encryption granularities also remains an interesting future direction.

\section{Acknowledgments}
This work is supported by NSF under Grants 1901137, 2106463, 2319962, 2019163, and 2126148.

\balance
\bibliographystyle{ACM-Reference-Format}
\bibliography{refs, susmit}

\end{document}